# A System for Nuclear Waste Transmutation Driven by Target-Distributed Accelerators

Anatoly Blanovsky
*Teacher Technology Center, 7850 Melrose Ave., Los Angeles, CA 90046, USA*

**Abstract-***A radioactive waste transmutation system based extensively on existing nuclear power technology is presented. By replacing the control rods with neutron sources, we could maintain good power distribution and perform long-lived waste burning in high flux subcritical reactors (HFSR). To increase neutron source intensity the HFSR is divided into two zones: a booster and a blanket. A neutron gate (absorber and moderator) imposed between two zones permits fast neutrons from the booster to flow to the blanket. Neutrons moving in the reverse direction are moderated and absorbed. The design is based on a small pressurized water reactor (PWR), fission electric cell (FEC), target-distributed accelerator (TDA) and power monitoring system with in-core gamma-ray detectors, now under development in several countries. The FEC is essentially a high-voltage power source that directly converts the kinetic energy of the fission fragments into electrical potential of about 2MV. The TDA, in which an FEC electric field compensates for lost beam energy in the target, offers a new approach to obtain large neutron fluxes.*

## INTRODUCTION

In the conventional reactor, nuclear energy (electrical in nature) is converted to thermal, then to mechanical and then to electrical energy. To achieve a high degree of burn-up, only fresh low enriched uranium fuel is used. Because surplus reactivity is large, a large amount of the burnable poison material has to be put in the conventional reactor at the expense of the neutron economy. A typical fuel assembly remains in the PWR for 3 years to a total burn-up of thirty thousand MWD/ton. The burn-up limitation is mainly because of criticality, but not due to radiation damage to the fuel elements. The burn-up for a total of sixty thousand MWD/ton is possible with conventional zircaloy clad fuel elements.

Several nations have programs to convert the PWR or fast breeder for waste burning. However, fast reactors have the high cost and long campaign. The thermal breeder is less expensive and safer than fast reactor. The concept is based on the use of a seed-blanket unit (SBU) pioneered at the Light Water Breeder Reactor. By lower the ratio of the volume of light water to the volume of fuel material, we can raise the average energy of neutrons to increase the conversion rate in the PWR.

The majority of the fission wastes have a half-life of less than one year. However, some fission waste products, as well as actinides, need long-term storage. The loss of about 99% of the fuel energy content and concern about the safety of the burial solution justify alternative studies. One such alternative is the HFSR with solid and liquid fuels.

## I. HFSR CONCEPT

The HFSR is mainly an epithermal breeder controlled by external neutron source intensity (source-dominated mode) or by variable feedback (self-amplifying mode). Managing the waste is significantly influenced by fission of the actinides. The most effective way of using the HFSR is to burn the actinides dissolved in water or in molten salt.

In this design, the plutonium-neptunium fuel circulates in the blanket and the booster; the americium-curium fuel circulates in the outer reflector and absorber zone. Use of a liquid actinide fuel permits transport of the delayed-neutron emitters from the blanket to the booster where they can provide additional neutrons or all the necessary excitation without an external neutron source[1].

With blanket and booster multiplication factors of k=0.95 and 0.98, respectively, an external neutron source rate of at least $10^{15}$ n/s is needed to drive the self-amplifying HFSR that produces 300MWt. Most of this power could be generated in the blanket that has actinide transmutation rate of about 100 kg/year.

Since Np-237 and Am-241 fission cross-sections are relatively low they would serve as absorbers in the low neutron flux of the dense absorber zone. In the high thermal neutron flux of the blanket, they would serve as a fuel. The effect of the fuel circulation throughout the core and cooling system on the HFSR reactivity is in the increasing role of delayed neutrons in the blanket.

## I.A. Actinide Transmutation in the HFSR

The HFSR module includes the distributed target, the separators, the booster and the blanket that consists of the SBU arranged in annular rings. It could be placed in the PWR pressure vessel, and a traditional system could be used to pump coolant to steam generators. The internal separators include well-shielded mini-tanks loaded with actinides from PWR spent fuel, and heat exchangers for heat transfer and some fission product deposition. In order to prevent the loss of neutrons, the reflectors surround the booster and the blanket (see Fig.1).

The modified SBU includes a central zone with several channels containing liquid actinide fuel. Preferably the actinides would be in the form of a suspension or salts dissolved as fluorides in water. A zone of vented depleted uranium or thorium fuel rods surrounds the central zone. By replacing light water with heavy water in the PWR fuel assemblies, we can shift the neutron spectrum to higher energies and efficiently convert U-238 into fissionable nuclei. Removing gas and volatile precursors of fission products with high thermal cross sections could reduce a neutron poison.

A preliminary analysis indicates that average thermal neutron fluxes of about $10.\text{sup}.15$ n/cm2s are achievable in the blanket with a power density of 300W/cm$^3$ and tight heavy water lattice. To achieve a significant actinide burning efficiency, the volume ratio of the heavy water moderator to fuel in the central region of the fuel assembly is set as high as possible. In the high thermal neutron flux, the number of neutrons required for the destruction of Np-237 and Am-241 by fission is substantially less than if decay allowed Np-238 and Am-242 to proceed farther up in the neutron capture chain[2].

In the central zone, the Pu-239 content will be much reduced because of the physical separation of U-238 and actinides. Plutonium with a high Pu-241 content that has an extremely high value of eta at epithermal energy and the delayed-neutron emitters are then continually transported into the booster. At a fissile fuel concentration of about 0.007 kg/l and depleted fuel concentration of 2.5 kg/l, the blanket would have k=0.95 up to maximum burn-up.

There is also about 10kg of actinide fuel in the mini-tanks outside of the core. It is continually fed into the core at the rate up to 300 g/day, and water with equilibrium concentration of actinides is removed to a temporary waste storage. As the primary purpose of the actinide feed is reactivity control, a flow control valve for each mini-tank provides preset flow rate. The actinide fractional feeding is optimized to flatten power distribution.

As the plutonium inventory in the blanket increases, the actinide feeding can be reduced. After a few fuel cycles with a solid fuel replacement, the next feed for actinide on-site transmutation is prepared by removing the cladding metal and uranium separation. For example, a Purex process can be used to separate neptunium and a Truex process to separate americium-curium from waste.

Long-lived radioactive waste transmutation could be based on a microwave separator, in which a feed material consisting of mixed isotopes is ionized by electron cyclotron resonance. The ionized material is then fed to an ion cyclotron resonance unit, which is tuned to preferentially excite the minor species.

At present, the cost of small power reactors is about $2 billion (the reactor component 33%, the turbine/generator 25% and other 42%). If the TDA cost were about $100 million, the HFSR cost would be equal or less than the cost of the present PWR, especially with the use of out-of-commission reactors.

## I.B. HFSR Control System

Feedback signals from the FEC arrays and calorimetric gamma-ray detectors could be used to control the power of the accelerator sections. These detectors were tested at the Armenian nuclear power plant. The power distribution and thermal state of the VVER-440 were computed every 20 seconds on the basis of 2K signals. The time-dependent power-to-signal conversion factor was determined from the previous values by using a simple recurrent formula. Off-line calculations were used for real-time synthesis of the signals into 3-D power distribution[3].

As the relaxation length of the power release is $L/(1-k)^{1/2}$, where L is the neutron migration length, the HFSR axial and radial distributions is similar to a commercial PWR. Photoneutrons produced in the blanket play an important part in the reactor control. Since the reactivity control can be basically performed electronically by using liquid fuel flow control valves, this simplifies core control and improves neutron economy. Conventional poison rods can be employed only for emergency shutdown of the reactor.

The HFSR control system will also include a digital reactivity meter for subcriticality monitoring and a TV satellite dish with a built-in GPS receiver to analyze the ultrasonic concept of the plant seismic reinforcement. To monitor crustal deformation of the earth, the system requires radio signals received by antennas on a long baseline. As such motions are preceded by an earthquake, a GPS-based earthquake prediction technique might be developed.

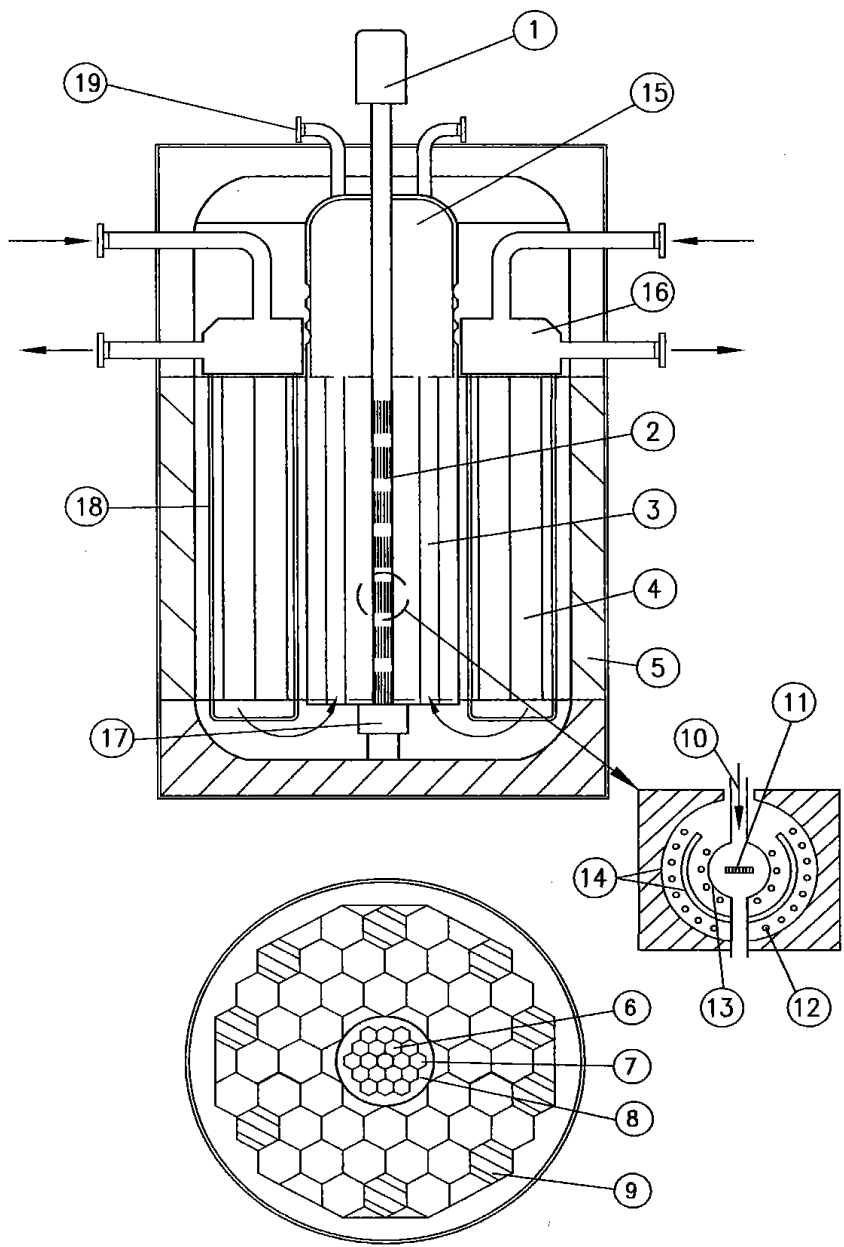

FIGURE 1. SCHEMATIC LAYOUT OF THE HFSR MODULE.

1. Accelerator; 2. Target; 3. Booster; 4. Blanket; 5. Reflector; 6. FEC; 7. Enriched Uranium; 8. Moderator; 9. Spent Fuel; 10. Beam; 11. Target; 12. Grid; 13. Cathode; 14. Anode; 15. Inner Vessel; 16. Separator; 17. Load; 18. Loop; 19. Vacuum pipe.

## II. PHOTONEUTRON TDA

Now, the most used photoneutron target is a thick tungsten layer followed by a beryllium layer. As the photon yield has maximum at depth of about 0.3-0.5R, where R is the range of electrons, the target made of a thin high-Z metal, such as tungsten, lead or tantalum, with heavy water cooling could be used. After impacting the thin target, the electrons are post-accelerated by the distributed transformer-rectifier circuits or the FEC.

Since an external electric field acts counter to the stopping power in the target or between the thin targets, the beam energy E is approximately constant. As a result, the neutron yield per particle is $Y(E)= d/L(E)$, where d is the thickness of the target. For n thin targets, in which the energy loss is regained by the acceleration of the particles between targets, the total neutron yield per particle is nY.

The Monte Carlo calculation of the total neutron yield per incident electron was made using MCNPX code. With 10 thin lead targets (0.2cm) in a thick lead tube and 30MV post acceleration, nY is about 0.02 at electron energy of 18MeV. With current 0.05A and total accelerating potential 48MeV the electron beam power is about 2.5MW.

Target heating should not be a problem since the main energy loss by electron beam is radiative. Photoneutrons are mostly generated in actinides, heavy water and beryllium surrounded photon production targets. The possible candidates of the target materials are actinide salt in heavy water as an electrolyte is a highly condensed plasma, and thorium oxide as it has high dielectric constant in optical range of frequency. Also, the TDA, in which an electric field compensates for lost beam energy in the gas D-T target, can be used.

The FEC array forms a high-voltage power source connected to the TDA. Each FEC has a tube-type thin fuel cathode with ribs and a quasi-spherical or cylindrical anode, nested in a hexagonal moderator. As this design has an extremely low fuel inventory, a liquid actinide fuel might be used. After the fuel fission, the more massive fission fragments can reach the anode and deposit their charges.

Since the fission fragments are naturally separated into two energy groups, a high efficiency and better electrical field distribution can be obtained through the use of a multistage collector (anode). With a two-stage collector, the second collector is made opaque. The first collector made of thin metal ribbons is essentially transparent to the incoming fragments but it is opaque to the fragments that are turned around. This technique is the most efficient (about 60%) because of its excellent space-charge handling ability[4].

The charge deposited by the electron beam in the target could be used to suppress the flow of secondary electrons across the gap between the FEC electrodes. As the stopping power is slightly increased, when the charge state of ions increases, the additional electric field across the cathode could increase the FEC electrical efficiency. The possible choices for the cathode materials are colored alkali halides to maximize the field strength in the cathode. Also, a relativistic magnetron driven by the FEC can be used as a microwave power sources for the TDA. The electron-screening effects that are controlled by the number of allowed energy states in medium and vacuum could play essential role in this design.

For a sphere with uniform density of sources S and macroscopic absorption cross-sections $\Sigma_a$, the average flux is $F \approx S/\pi*(\Sigma_a+DB^2)$ and leakage probability $P=B^2L^2/(1+ B^2L^2)$. Here the buckling is $B=\pi/R$ and the diffusion mean free path is $L^2=D/\Sigma_a$. Feedback is important in this design. As a liquid fuel transports a fraction of the delayed-neutrons $\alpha$ from the blanket, it could essentially improve the FEC performance. If $D=0.7$cm, $\Sigma_a=10^{-3}$cm$^{-1}$, $R=15$cm, $L=25$cm and $\alpha=0.25$, we have $S \approx 10^{12}$ n/s*cm$^3$ and $F \approx 7*10^{12}$ n/s*cm$^2$. The fission fragment current is about 0.5mA/g for the FEC fuel thickness of 1.5μm[5].

## III. CONCLUSION

Although low energy accelerators are mainly seen as support equipment, recent advances in neutron generator technology have made it possible to use them for energy generation systems. Different neutron producing reactions (D-T, p-Li, γ-n, etc.) can be used. The initial HFSR design could be based on the intense D-T or resonance neutron sources and circulating-fuel reactor technology. The final design would require analysis of the relativistic effects in the electron TDA and in the FEC.